\documentclass[12pt,a4paper,DIV12]{scrartcl}

\usepackage[utf8]{inputenc} 
\usepackage[T1]{fontenc}
\usepackage{lmodern}
\usepackage[british]{babel}

\usepackage{hyperref}
\usepackage{grffile}

\usepackage{graphicx}
\usepackage{grffile}
\usepackage{color}

\usepackage{amssymb}
\usepackage{amsmath}
\usepackage{latexsym}
\usepackage{amsthm}
\usepackage{eucal}

\usepackage{bm} 
\usepackage{slashed}

\usepackage{float} 

\newcommand {\beq}{\begin{equation}}
\newcommand {\eeq}{\end{equation}}

\newcommand {\bea}{\begin{eqnarray}}
\newcommand {\eea}{\end{eqnarray}}

\newcommand*{\email}[1]{\href{mailto:#1}{#1}} 

\title{%
  {\vspace{-20mm}\normalsize
  \hfill\parbox[b][30mm][t]{35mm}{\textmd{MS-TP-17-07}}}\\[-18mm]
Improved thermodynamics of SU(2) gauge theory
\vspace*{2mm}}
\author{%
Pietro Giudice\\
\textit{\large University of M\"unster, Institute for Theoretical Physics}\\
\textit{\large Wilhelm-Klemm-Str.~9, D-48149 M\"unster, Germany}\\
\textit{\large E-mail: \email{p.giudice@uni-muenster.de}}\\[5mm]
Stefano Piemonte\\
\textit{\large University of Regensburg, Institute for Theoretical Physics}\\
\textit{\large Universit\"atsstr.~31, D-93040 Regensburg, Germany}\\
\textit{\large E-mail: \email{stefano.piemonte@ur.de}}
\vspace*{5mm}}

\date{\today}

\begin{document}
\maketitle


\begin{abstract}
\noindent
\textbf{\textsf{Abstract:}}
In this work we present the results of our investigation about
the thermodynamics of SU(2) gauge theory.
We employ a Symanzik improved action to reduce strongly the discretisations
effects, and we use the scaling relations to take into account
the finite volume effects close to the critical temperature.
We determine the $\beta$-function for this particular theory and we use it 
in the determination of different thermodynamic observables.
Finally we compare our results with previous works where only the standard
Wilson action was considered. We confirm the relevance of
using the improved action to access easily the correct continuum thermodynamics 
of the theory.

\end{abstract}

\newpage

\section{Introduction}

Asymptotic freedom and confinement are two crucial properties of QCD. Confinement implies that the fundamental degrees of freedom of the theory, namely quarks and gluons, cannot be found as isolated particles in nature but exist only in complex bound states under normal conditions. The confinement properties of QCD-like theories are very well described in terms of a flux tube arising between quark-antiquark static charges. The vacuum quantum fluctuations of the flux tube are expected to be described by non-critical string models, and, for pure gauge theories without light quarks, string breaking does not occur and the accuracy of the predictions has been verified by many lattice Monte Carlo simulations. Despite the absence of dynamical quarks, the bound spectrum of pure gauge theories is still non-trivial due to the emergence of composite particles from the strong interactions between gluons, the so-called glueballs. There have been much effort in the past years to determine their properties at zero temperature,
see Refs.~\cite{Chowdhury:2014mra, Chen:2005mg, Loan:2006gm}, and also at non-zero temperature, see 
Refs.~\cite{Caselle:2013qpa, Meng:2009hh, Loan:2008nz, Ishii:2002ww}, although, an unambiguous experimental confirmation of their existence is still missing.

The fundamental nature of strong interactions is however quite different at very high temperature, where QCD behaves as a gas of free quarks and gluons due to asymptotic freedom. Understanding what happens to QCD for intermediate temperatures, near the deconfinement phase transition, is therefore the main reason for 
studying the thermodynamics of gauge theories. The chromodynamic flux tube is expected to survive below the critical temperature of quark deconfinement; various models have been developed to include thermal fluctuations to the QCD string, see for instance Refs.~\cite{Cea:2015wjd,Bicudo:2017uyy,Caselle:2011vk}.

Quark-gluon plasma (QGP) is the phase of matter that can be probed experimentally by particle accelerators, such as RHIC and LHC, occurring at temperatures higher than $\approx 200$ MeV. The properties of QGP even at quite large temperatures are compatible with those of a strongly interacting plasma that can be viewed as a perfect liquid, where color charges have long range interactions \cite{Heinz,Shuryak}. Because the success of the hydrodynamical description of high-energy heavy ion reactions, it is of great interest to compute the shear and bulk viscosities of the quark-gluon plasma. Because of the strongly interacting nature of the QGP, weak coupling perturbation theory is not able to capture the full thermal behavior of QCD. Lattice Monte Carlo simulations can provide a non-perturbative insight to the thermodynamics of the quark-gluon plasma, but still today it is not possible to compute the shear and bulk viscosities in full QCD and even in pure gluodynamics is an  extremely complicated task, see Refs.~\cite{Astrakhantsev:2017nrs,Pasztor:2016wxq, Astrakhantsev:2015jta}.

The properties of pure gauge theories at non-zero temperature have been intensively investigated based on the idea that it is possible to describe the 
thermodynamics of $N_c=3$ QCD as a limiting case of a $1/N_c$ expansion~\cite{Panero:2009tv,Bringoltz:2005rr,Datta:2010sq}. In particular, Feynman diagrams including quark lines give only a subleading contributions at large $N_c$.

In the same line of research, there has been several predictions for the behavior of the quark-gluon plasma after the deconfinement phase transition in the context of the AdS/CFT correspondence. Interesting comparisons have been made in 
Ref.~\cite{Panero:2009tv} with the so-called improved holographic QCD model,  
proposed in Ref.~\cite{Gursoy:2008za}.

In Ref.~\cite{Castorina:2011ra} SU(N$_c$) gauge theories are instead compared  with the quasi-particle approach.
It turns out that it gives a very good description of the interaction measure, {\itshape i.e.} of the trace anomaly, and of the thermodynamical
quantities.

All works, based on the lattice approach, which want to explore the large $N_c$-limit, 
are based on simulations with $N_c \geq 3$. 
The reason is that, while the deconfinement transition
for $N_c > 3$ is a first order phase transition and for $N_c=3$ is 
a weak first order one, 
the case with $N_c=2$ is characterised by a second order phase transition.
It is therefore expected that the models which describe the theories 
for $N_c \geq 3$ cannot describe the case with $N_c=2$ because 
this theory is qualitatively and quantitatively different.

Unfortunately, because simulations are missing, we do not know
so much about its properties and we do not know how much SU(2) pure gauge is 
really different with respect to  $N_c \geq 3$.
The last systematic studies, concerning the thermodynamic properties
of SU(2) gauge theory, go back to the beginning of the 
nineties, see {\itshape e.g.} Refs.~\cite{Engels:1988ph, Engels:1990vr}
and Ref.~\cite{Engels:1994xj} where only the energy density was considered.
We believe that it is timely to perform such a study and
this paper is devoted to this task.

The simulations in this work have been done using the Symanzik
improved action with 6-links plaquette. This is an important
aspect of our work. It is well known, see Ref.~\cite{Karsch:1996ja}, 
that the standard Wilson action,
with temporal extension $N_\tau=4$, leads to almost $50\%$ of corrections
due to finite cut-off effects, but using our improved action this
is reduced below $2\%$. We do not need therefore to simulate the theory
with high values of $N_\tau$, making the entire work much cheaper.
For example, in Ref.~\cite{Borsanyi:2012ve} the authors simulated SU(3) 
gauge theory with $N_\tau$ in the range $[5,8]$, which is by far much
more expensive. From a numerical point of view, the main difficulties come from
the finite volume effects close to the deconfinement transition
which is due to the second order phase transition.

Some thermodynamic quantities need the evaluation of the $\beta$-function;
this task is performed in Section~\ref{sec:betafunct}.
In Section~\ref{sec:themodynamics}, we measure a number of thermodynamic 
observables and we plot them. We compare our results with other works
in Section~\ref{sec:comparison} and finally we draw our conclusions
in Section~\ref{sec:conclusion}.


\section{Scale setting and the determination of the \texorpdfstring{$\beta$-function}{beta-function} }
\label{sec:betafunct}

The determination of the Callan-Symanzik $\beta$-function is relevant to set the scale, {\itshape i.e.} the physical temperature realised in our simulations. It is also important to determine how the physical volumes of our lattice changes when the bare coupling $g$ changes, that is the starting point to derive the trace of the energy-momentum tensor at non-zero temperature. While there are different papers where the $\beta$-function has been determined for the case of the standard Wilson action, see {\itshape e.g.} Refs.~\cite{Engels:1994xj, Pennanen:1997qm, Bali:1994de}, there are no works, to best of our knowledge, in the case of the action used in this work, therefore we must proceed to a separate calculation.

The physical measure of the lattice spacing $a$ as a function of the inverse coupling $\beta=4/g^2$ is performed by determining how a specified observable depends on $\beta$.
In literature different observables have been considered, previous works include the plaquette, 
see Refs.~\cite{Datta:2010sq, Datta:2009jn}, and the string 
tension, see Ref.~\cite{Lucini:2005vg}. In this work we consider three different observables: 
the critical temperature $T_c$ of the deconfinement phase transition, given by the critical coupling $\beta_c$ for different values of the lattice extent in the temporal direction  $N_\tau$, the scale parameter $w_0$, see Ref.~\cite{Borsanyi:2012zs},
and the scale parameter $t_0$, see Ref.~\cite{Luscher:2010iy}.

The $\beta$-function has been fitted starting from the expected scaling of physical observables near the continuum limit\footnote{In this paper 
we follow the convention that a hat above an observable, like in $\hat{O}$, 
means dimensionless quantity.}:
\bea
N_\tau(g^2)&=&\frac{1}{a(g^2) T_c}         \ , \label{ntauvsa} \\
\hat{w}_0(g^2)&=&\frac{w_0}{a(g^2)}   \ , \label{w0vsa} \\ 
\hat{t}_0(g^2)&=&\frac{t_0}{a^2(g^2)} \ . \label{t0vsa} 
\eea
The running of the lattice spacing $a$ as a function of $g$ is provided by the scaling function $F(g^2)$ up to corrections $A(g^2)$, which takes into account the lattice artifacts~\cite{Allton:1996kr}
\beq
a^{-1}= \frac{\Lambda_{\textrm{Lat}}}{F(g^2)} A(g^2) \ .
\label{eq:inversea}
\eeq
The scaling function $F(g^2)$ is given by the product of two terms:
\beq
F(g^2)=f_{PT}(g^2) \ \lambda(g^2) \ ;
\eeq
the first one is the result of the integration of the two-loop scheme-independent weak
expansion of the $\beta$-function
\beq
f_{PT}(g^2)=\exp{\left( -\frac{b_1}{2 b_0^2} \ln{(b_0 g^2) 
-\frac{1}{2 b_0 g^2} } \right)} \ ,
\label{eq:fperturb}
\eeq
while the term $\lambda(g^2)$ takes into account the terms of higher order of perturbation theory.
This term has been parametrised in different ways in literature; we have
considered two of them. The first one, see Ref.~\cite{Engels:1994xj}:
\beq
\lambda(g^2)=\exp{\left[ 
\frac{1}{2 b_0^2}(c_1 g^2 + c_2 g^4 + c_3 g^6 + \ldots ) 
\right]} \ ,
\eeq
and the second one, see Ref.\cite{Allton:1996kr}:
\beq
\lambda^{\prime}(g^2)=1+d_2 g^2 + d_3 g^4 + d_4 g^6 + \ldots
\eeq
We have verified that the second method gives worst results, in particular when 
the correction $A(g^2)$ is considered. Therefore in the following we will 
show only the results obtained with the functional form $\lambda(g^2)$.

The function $A(g^2)$ accounts for scaling violations far away from the continuum limit driven by the running of irrelevant lattice operators at non zero lattice spacing $a$, see Refs.~\cite{Allton:1996kr, Trivini:2005ag}. The form of $A(g^2)$, 
\beq
A(g^2)=1 - X_{n,\nu} \ g^\nu \ \left( \frac{f_{PT}(g^2) }{f_{PT}(1)} \right)^n
- Y_{n^\prime,\nu^\prime} \ g^{\nu^\prime} \ 
\left( \frac{f_{PT}(g^2)}{f_{PT}(1)}\right)^{n^\prime} \ ,
\eeq
is specified in terms of two even integer numbers $\nu$ and $\nu^\prime$, 
because we require that $a$ is an even function of $g$.  The term containing $X_{n,\nu}$ takes into 
account the leading correction in $a$; the term $Y_{n^\prime,\nu^\prime}$ the
next-to-leading one. Each term has been normalised so that $X_{n,\nu}$ and
 $Y_{n^\prime,\nu^\prime}$ describe the fractional amount of scaling correction
at a standard value of $g=\sqrt{2}$, corresponding to $\beta=4/g^2=2$.

The $\beta$-function $\beta_f$ can be expressed as a function of $\beta=4/g^2$:
\beq
\beta_f = -a \frac{\partial g}{\partial a} = 
\frac{1}{\beta^{3/2}} \  \frac{\partial \beta}{\partial \log(a)} \ ,
\eeq
where the term ${\partial \beta}/{\partial \log(a)}$ can be easily determined
using Eqs.~(\ref{ntauvsa}),~(\ref{w0vsa}),~(\ref{t0vsa}):
\bea
\frac{\partial \beta}{\partial \log(a)} &=& - \left( \frac{1}{N_\tau} \frac{\partial N_\tau}{\partial \beta} \right)^{-1} \ , \\
\frac{\partial \beta}{\partial \log(a)} &=& - \left( \frac{1}{\hat{w}_0} \frac{\partial \hat{w}_0}{\partial \beta}  \right)^{-1} \ , \\
\frac{\partial \beta}{\partial \log(a)} &=& -2 \left( \frac{1}{\hat{t}_0} \frac{\partial \hat{t}_0}{\partial \beta}  \right)^{-1} \ .
\eea
Starting from these relations we can determine three different definitions of the lattice $\beta$-function; 
in the following sections we present each definition and the resulting scale in detail.

\subsection{Fitting the critical \texorpdfstring{$\beta$}{beta}}

The $\beta$-function can be determined from the value of the critical bare gauge coupling $g$ where the deconfinement phase transition occurs at different values of the lattice temporal extension $N_\tau$. To this end, we can exploit existing calculations presented in literature, see Ref.~\cite{Cella:1993ic} and summarised in Table~\ref{tab:ntvsbetacrit}.

\begin{table}[htbp]
  \begin{center}
    \begin{tabular}{|l|l|}
      \hline
       $N_{\tau}$    &  $\beta_c$  \\
      \hline
      3 &   1.59624(13) \\
      4 &   1.699(1)    \\
      6 &   1.8287(11)  \\
      7 &   1.8747(30)  \\
      8 &   1.920(5)    \\
      \hline
    \end{tabular}
  \end{center}
  \caption{Critical value of $\beta_c=4/g^2$ for different values of $N_\tau$ from Ref.~\cite{Cella:1993ic}.
Note that the value at $N_\tau=5$ has not been considered because it was clearly too far from 
the interpolating function, perhaps the error cited in Ref.~\cite{Cella:1993ic} has been underestimated.}
  \label{tab:ntvsbetacrit}
\end{table}

The points $(\beta_c, N_\tau)$ are fitted using Eq.~(\ref{ntauvsa}), where the lattice spacing is given by 
Eq.~(\ref{eq:inversea})
\beq
N_\tau=\frac{\Lambda_{\textrm{Lat}} /T_c}{F(g^2)} A(g^2) = \frac{\Omega^\prime}{F(\beta)} A(\beta) \,,
\eeq
and $\Omega^\prime$ is the dimensionless ratio $\Lambda_{\textrm{Lat}}/T_c$.

Note that for a given integer value of $N_\tau$, the deconfinement phase transition is located by looking for the maximum of the Polyakov loop susceptibilities as a function of $\beta$ and not vice-versa. Therefore in this fitting procedure the error appears to be on the abscissa, on $\beta$, and there is no
error on the ordinate, on $N_\tau$. In general, when in this work we have to combine errors
in both dimensions, let us say $\sigma_x$ and $\sigma_y$, we use the approach explained 
in Ref.~\cite{D'Agostini:2005pk}, {\itshape i.e.} we combine the two errors as $\sqrt{ s^2 \sigma_x^2 + \sigma_y^2}$,
where $s$ is the slope of the curve in the point we are considering.
In Table~\ref{tab:fit_many_Nt} we report the value of $\chi^2/$d.o.f. for different fits of the data 
presented in Table~\ref{tab:ntvsbetacrit}.

\begin{figure}[H]
  \centering
  \includegraphics[width=0.75\textwidth]{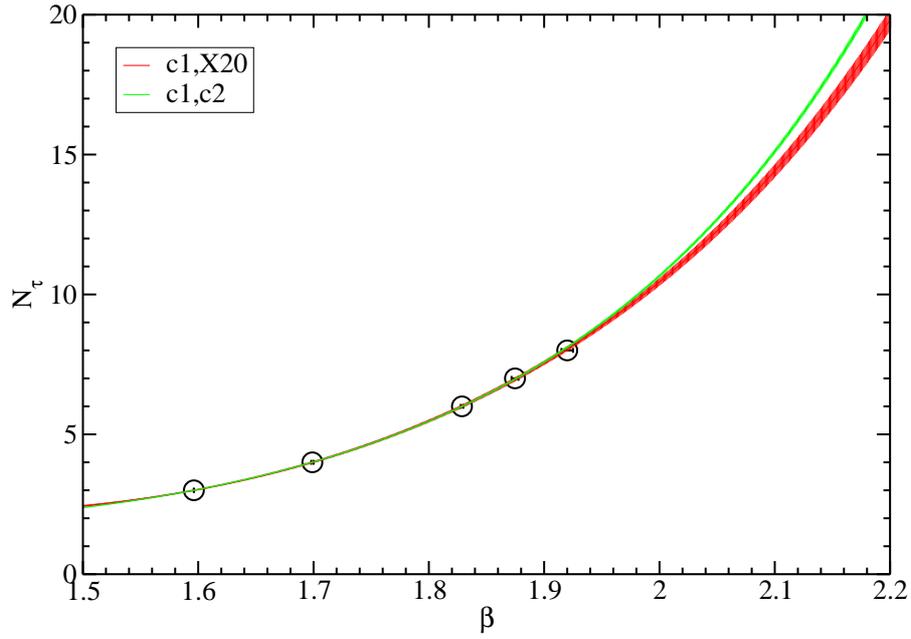}
  \caption{Interpolation of $N_\tau$ versus $\beta$ using ``fit2'' and ``fit3'' of Table~\ref{tab:fit_many_Nt}.}
  \label{fig:Nt_versus_beta}
\end{figure}

\begin{table}[htbp]
  \begin{center}
    \begin{tabular}{|l|l|l|l|l|l|}
      \hline
      & fit parameters & $\chi^2/\mbox{d.o.f.}$  \\
      \hline
      fit1 & $\Omega^\prime,c_1$ & 19.30   \\
      \hline
      fit2 & $\Omega^\prime,c_1,c_2$ & 0.45   \\
      \hline
      fit3 & $\Omega^\prime,c_1,X_{2,0} $ & 0.28  \\
      \hline
      fit4 & $\Omega^\prime,c_1,c_2,c_3$ & unstable   \\
      \hline
      fit5 & $\Omega^\prime,c_1,c_2,X_{2,0} $ & unstable  \\
      \hline
    \end{tabular}
  \end{center}
  \caption{Summary of $\chi^2/\mbox{d.o.f.}$ for different types of fits of $N_\tau$ vs $\beta_c$. 
    $\Omega^\prime=\Lambda_{\textrm{Lat}}/T_c$. }
  \label{tab:fit_many_Nt}
\end{table}
According to this table, the best two fits, with the smaller $\chi^2/\mbox{d.o.f.}$, 
are those labelled with ``fit2'' and ``fit3''.  The final result is presented in 
Figure~\ref{fig:Nt_versus_beta} where the error is the statistical one.

\begin{figure}[H]
  \centering
  \includegraphics[width=0.75\textwidth]{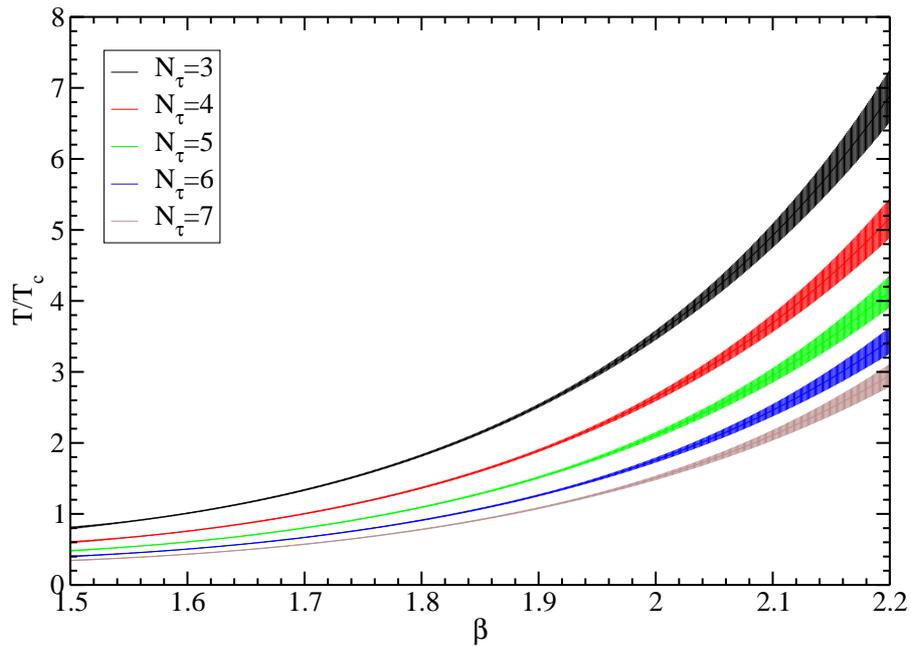}
  \caption{$T/T_c$ vs $\beta$ for different values of $N_\tau$. The error band has 
    been determined considering the sum of statistical and systematic errors due 
    to the difference of the two interpolations of Figure~\ref{fig:Nt_versus_beta} and
    as central value the average of the interpolations.}
  \label{fig:TvsTc_versus_beta}
\end{figure}

In Figure~\ref{fig:TvsTc_versus_beta} we plot the ratio $T/T_c$ versus
$\beta$ for different values of $N_\tau$ as determined from data of 
Figure~\ref{fig:Nt_versus_beta}. Here the value is given as average 
of the previous two best fits and the error takes into account the
statistical error of the two fits and the systematic error given by the  
difference between the two curves.
The values presented in this plot are the ones used in the remain part 
of this work every time we need a correspondence between $\beta$
and $T/T_c$.

\subsection{Fitting the scale parameter \texorpdfstring{$w_0$}{w0} and \texorpdfstring{$t_0$}{t0}}

The numerical integration of the flow equations, defined from the functional derivative of the Symanzik gauge action with respect to the gauge-link variables, is performed using fourth order Runge-Kutta integrator with a discretisation of the flow time equal to $\delta t = 0.01$. We have measured the energy density, defined to be equal to the traceless anti-hermitian part of the clover plaquette, every ten integration steps. The scale parameter $w_0$ is defined as the square root of the flow time $t$, solution of the implicit equation
\beq\label{w0def}
t \frac{d}{dt} t^2 \langle E(t) \rangle = u \,.
\eeq 
We have used two different values for the reference value $u$ (0.2 and 0.3). The observable $t_0$, defined as the flow time $t$ where the equation
\beq
t^2 \langle E(t) \rangle = u \,,
\eeq 
is fulfilled, is expected to be affected by larger discretisation effects, see Ref.~\cite{Borsanyi:2012zs}, that will appear as scaling violations in our fitting procedure. To compute the logarithmic derivative in Eq.~(\ref{w0def}), we have performed a polynomial fit of the expectation value of the flow observable $\langle E(t) \rangle$. Given that the flowed energy density is strongly correlated for flow times close to each other, we estimate the statistical error using the bootstrapping method, performing therefore a fit on each bootstrapping sample. In our tables we quote only the statistical error, without the systematic error coming from various possible fitting intervals and degrees of the interpolating polynomial.

The values of the scale $w_0(\beta)/a$ and $t_0(\beta)/a^2$, that we have measured, can be found in Table~\ref{tab:w0values}. In the last column appears also the value of 
the ``residual'' non-zero temperature of the system determined by the ratio of the critical $N_\tau$ plotted in Figure~\ref{fig:Nt_versus_beta} and the size of the 
four-dimensional hypercube $\hat{L}$. 

Note that, for $\beta=1.8$, $\beta=1.825$ and $\beta=1.85$, we measured the scales for three different volumes, up to $T /T_c \approx 0.36$,
and the difference never exceeds three standard deviations. Since $T /T_c \approx 0.36$ is also the maximum ``residual'' temperature corresponding to the larger
value of $\beta$ that we have used in our simulations, {\itshape i.e.} $\beta=2.025$, we can safely assume that the finite volume effects are under control
in the entire range of $\beta$.
Moreover, as we will show in Sec.~\ref{sub:simulations}, within the precision of our measurements, with volumes ranging from $\hat{L}=24$ to $\hat{L}=56$, 
and with a range of $\beta$ corresponding to a “residual” non-zero temperature below $T /T_c \approx 0.45$, also the spatial plaquette
is not affected by finite volume effects.

About 1000 configurations were discarded for thermalisation; 
200 configurations were generated with $\hat{L}=32$ and 350 with $\hat{L}=24$.
The measurements are separated by 30 iterations of combined Cabibbo-Marinari 
heatbath and overrelaxation sweeps. 

\begin{table}[htbp]
  \begin{center}
    \begin{tabular}{|l|l|l|l|l|l|l|}
      \hline
      $\beta$ & $\hat{w}_0^{u=0.2}$ &$\hat{w}_0^{u=0.3}$  & $\hat{t}_0^{u=0.2}$ &$\hat{t}_0^{u=0.3}$ & $\hat{L}$ & $T/T_c$  \\
      \hline
      1.550 & 0.770038(96) & 0.91289(15) & 0.71680(12)   &  0.9937(2)   & 24   & 0.11    \\
      \hline
      1.575 & 0.80416(10)  & 0.95377(17) & 0.77452(14)   &  1.07711(25) & 24   & 0.12     \\
      \hline
      1.600 & 0.84354(13)  & 0.99848(19) & 0.84243(17)   &  1.17431(29) & 24   & 0.13     \\
      \hline
      1.625 & 0.88891(20)  & 1.04996(27) & 0.92503(30)   &  1.29238(50) & 24   & 0.13     \\
      \hline
      1.650 & 0.94101(23)  & 1.10899(25) & 1.02526(38)   &  1.43549(56) & 24   & 0.14     \\
      \hline
      1.675 & 0.99824(27)  & 1.17329(40) & 1.14359(44)   &  1.60305(76) & 24   & 0.15     \\
      \hline
      1.700 & 1.06118(52)  & 1.24409(64) & 1.28394(89)   &  1.8007(14)  & 24   & 0.17     \\
      \hline
      1.725 & 1.13147(61)  & 1.32190(76) & 1.4517(12)    &  2.0350(19)  & 24   & 0.18     \\
      \hline 
      1.750 & 1.21446(99)  & 1.4159(11)  &  1.660(2)     &  2.3297(31)  & 24   & 0.19     \\
      \hline
      1.775 & 1.2995(16)   & 1.5105(18)  &  1.8927(37)   &  2.6546(56)  & 24   & 0.21     \\
      \hline
      1.800 & 1.3937(20)   & 1.6166(28)  &  2.1692(42)   &  3.0416(72)  & 18   & 0.30     \\
      \hline
      1.800 & 1.3991(14)   & 1.6232(18)  &  2.1843(34)   &  3.0638(58)  & 24   & 0.23     \\
      \hline
      1.800 & 1.3954(5)    & 1.61884(82) & 2.1754(13)    &  3.0506(22)  & 32   & 0.17     \\
      \hline
      1.825 & 1.5039(34)   & 1.7397(46)  & 2.5183(86)    &  3.528(14)   & 18   & 0.32     \\
      \hline
      1.825 & 1.4995(23)   & 1.7360(26)  & 2.5033(63)    &  3.5101(90)  & 24   & 0.24     \\
      \hline
      1.825 & 1.4993(10)   & 1.7366(13)  & 2.5006(24)    &  3.5080(39)  & 32   & 0.18     \\
      \hline
      1.850 & 1.6061(39)   & 1.8548(44)  &  2.871(11)    &  4.019(17)   & 18   & 0.36     \\
      \hline
      1.850 & 1.6186(47)   & 1.8714(54)  &  2.903(13)    &  4.072(19)   & 24   & 0.27     \\
      \hline
      1.850 & 1.6088(12)   & 1.8593(15)  &  2.8762(40)   &  4.0299(58)  & 32   & 0.20     \\
      \hline
      1.875 & 1.7317(14)   & 2.0011(17)  &  3.3147(41)   &  4.6529(64)  & 32   & 0.22     \\ 
      \hline
      1.900 & 1.8706(22)   & 2.1582(32)  &  3.8585(71)   &  5.415(12)   & 32   & 0.24     \\
      \hline
      1.925 & 2.0248(27)   & 2.3331(39)  & 4.496(11)     &  6.319(18)   & 32   & 0.26     \\
      \hline
      1.950 & 2.1767(32)   & 2.5106(36)  & 5.178(15)     &  7.291(21)   & 32   & 0.28     \\
      \hline
      1.975 & 2.326(11)    & 2.677(13)   & 5.909(46)     &  8.308(71)   & 32   & 0.30     \\
      \hline
      2.000 & 2.5155(98)   & 2.897(13)   &  6.872(40)    &   9.689(68)  & 32   & 0.33     \\
      \hline
      2.025 & 2.693(13)    & 3.099(14)   &  7.877(68)    &   11.074(63) & 32   & 0.36     \\
      \hline

    \end{tabular}
  \end{center}
  \caption{Summary of the values of $\hat{w}_0$ and $\hat{t}_0$ used to determine the 
$\beta$-function. The quoted error is only statistical and does not include systematic errors arising from different choices of the fit of the flow. $\hat{L}$ is the size of the hypercube used. $T/T_c$ is
the ``residual'' non-zero temperature of the system determined from Figure~\ref{fig:Nt_versus_beta}.
}
  \label{tab:w0values}
\end{table}

\begin{table}[htbp]
  \begin{center}
    \begin{tabular}{|l|l|l|l|l|l|}
      \hline
      & fit parameters & u=0.2 & u=0.3 \\
      \hline
      fit1 & $\Omega,c_1,c_2$ & 30.69 & 8.25  \\
      \hline
      fit2 & $\Omega,c_1,c_2,c_3$ & 5.23 & 7.77  \\
      \hline
      fit3 & $\Omega,c_1,c_2,c_3,c_4$ & 5.88 & 7.43   \\
      \hline
      fit4 & $\Omega,c_1,c_2,c_3,c_4,c_5$ & 5.39 & 6.25   \\
      \hline
      fit5 & $\Omega,c_1,X_{2,0}$ & 5.86 & --   \\
      \hline
      fit6 & $\Omega,c_1,c_2,X_{2,0}$ & 4.97 & --   \\
      \hline
      fit7 & $\Omega,c_1,c_2,c_3,X_{2,0}$ & 5.04 & --   \\
      \hline
      fit8 & $\Omega,c_1,c_2,c_3,c_4,X_{2,0}$ & 5.39 & --   \\
      \hline
    \end{tabular}
  \end{center}
  \caption{Summary of $\chi^2/\mbox{d.o.f.}$ for different types of fits of $\hat{w}_0$ vs $\beta$. 
    Here $\Omega=w_0 \Lambda_{\textrm{Lat}}$.}
  \label{tab:fit_many_w0}
\end{table}

The measured data $(\beta,w_0(\beta)/a)$ are fitted using Eq.~(\ref{w0vsa}), where the lattice spacing is 
given by Eq.~(\ref{eq:inversea}) and $\Omega=w_0 \Lambda_{\textrm{Lat}}$:
\beq
\hat{w}_0=\frac{w_0 \Lambda_{\textrm{Lat}}}{F(g^2)} A(g^2) = \frac{\Omega}{F(\beta)} A(\beta) \ .
\eeq
Similarly, using Eq.~(\ref{t0vsa}), we fit the scale $t_0(\beta)/a^2$ as
\beq
\hat{t}_0=\frac{t_0 \Lambda^2_{\textrm{Lat}}}{F^2} A^2 = \frac{\Omega^{\prime\prime}}{F^2(\beta)} A^2(\beta) \ ,
\eeq
where $\Omega^{\prime\prime}=t_0 \Lambda^2_{\textrm{Lat}}$. 

In Table~\ref{tab:fit_many_w0} we report the value of $\chi^2/$d.o.f. for different fits of the scale $w_0$ presented in 
Table~\ref{tab:w0values}. Comparing $u=0.2$ and $u=0.3$ data, we see that we have a better fit in the first case (for more 
than three fitting parameters). For $u=0.3$ and from ``fit5'' to ``fit8'' was not possible to fit the data because of numerical
instabilities. As discussed in Ref.~\cite{Bergner:2014ska} the value of $u$ cannot be too large otherwise 
the results obtained by the gradient flow can be negatively affected by finite volume effects and by large Monte Carlo autocorrelations. 
In the following we consider therefore only data obtained with $u=0.2$. 

\begin{table}[htbp]
  \begin{center}
    \begin{tabular}{|l|l|l|l|l|l|}
      \hline
      & fit parameters & u=0.2  \\
      \hline
      fit1 & $\Omega^{\prime\prime},c_1,c_2$ & 45  \\
      \hline
      fit2 & $\Omega^{\prime\prime},c_1,c_2,c_3$ & 24.1 \\
      \hline
      fit3 & $\Omega^{\prime\prime},c_1,c_2,c_3,c_4$ & 24.0   \\
      \hline
      fit4 & $\Omega^{\prime\prime},c_1,c_2,c_3,c_4,c_5$ & 15.3   \\
      \hline
      fit5 & $\Omega^{\prime\prime},c_1,c_2,X_{2,0}$ & 17.0   \\
      \hline
      fit6 & $\Omega^{\prime\prime},c_1,c_2,c_3,X_{2,0}$ & 10.3   \\
      \hline
      fit7 & $\Omega^{\prime\prime},c_1,c_2,c_3,c_4,X_{2,0}$ & 11.1   \\
      \hline
      fit8 & $\Omega^{\prime\prime},c_1,X_{2,2},Y_{4,0} $ & 308.2   \\
      \hline
      fit9 & $\Omega^{\prime\prime},c_1,c_2,X_{2,0},Y_{4,0} $ & 4.94   \\
      \hline
      fit10 & $\Omega^{\prime\prime},c_1,c_2,c_3,X_{2,0},Y_{4,0} $ & 5.37   \\
      \hline
      fit11 & $\Omega^{\prime\prime},c_1,c_2,c_3,c_4,X_{2,0},Y_{4,0} $ & 5.74   \\
      \hline
    \end{tabular}
  \end{center}
  \caption{Summary of $\chi^2/\mbox{d.o.f.}$ for different types of fits of $\hat{t}_0$ vs $\beta$.
    Here $\Omega^{\prime\prime}=t_0 \Lambda_{\textrm{Lat}}^2$.}
  \label{tab:fit_many_t0}
\end{table}
In the case of the scale $t_0$, we need to take into account the discretisation effects including the second order correction, {\itshape i.e.} 
taking into account also the coefficient $Y_{n,\nu}$, as it is evident from the results of the various fits in Table~\ref{tab:fit_many_t0}.

However we have to note that the value of $\chi^2/$d.o.f. is not of the order of one, but the best we could get is $\chi^2/\mbox{d.o.f.} \sim 5$. 
Such a large deviation from the asymptotic scaling is a typical situation that occurs when fits of non-perturbative results, using lattice perturbation theory, are considered. 
The $\chi^2/$d.o.f. could be improved if more fitting parameters were included, but in our case, given the non-linear form of our fitting function, such a method provides an unstable minimum of $\chi^2$. An another reason for the large $\chi^2/$d.o.f. is that the integrated energy of the flow is estimated rather precisely, so that the statistical error is comparable to the systematic error coming from finite volume effects and the various possible fitting range and fitting polynomials of the flow observable $\langle E(t) \rangle$. However,  we are not concerned of a such large $\chi^2/$d.o.f., since the largest systematic errors are coming rather from scaling violations as lattice artefacts, {\itshape i.e.} when different observables are used to set the scale. Our final $\beta$-function is a combination of three different definitions, see Sec.~\ref{sec:finalbeta}, and the final systematic error is large enough to accommodate any possible mismatch of our fits from the pertubative scaling.

\begin{figure}[H]
  \centering
  \hspace{1.5cm}
  \includegraphics[width=0.75\textwidth]{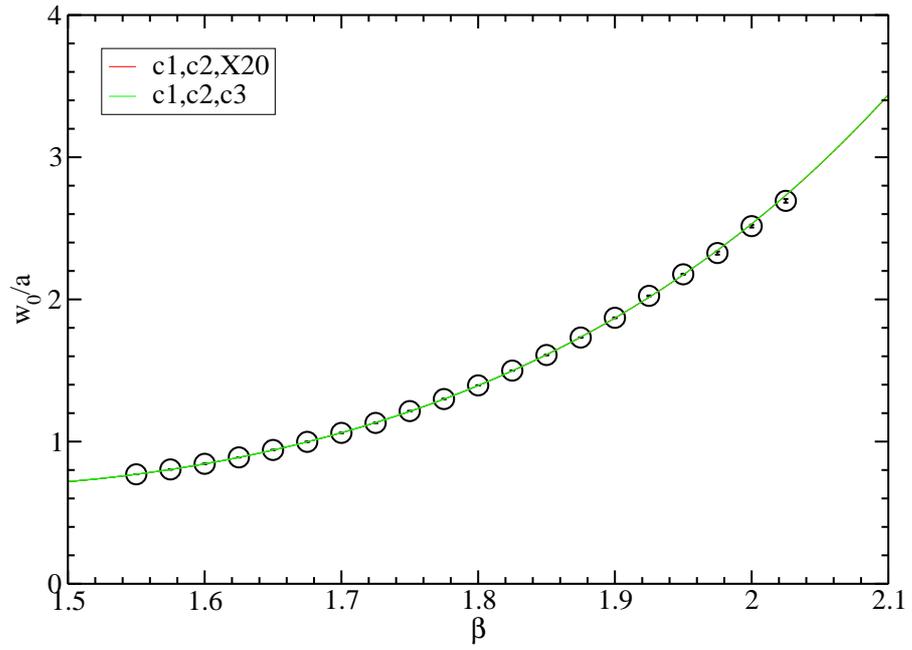}
  \caption{Interpolation of $w_0$ versus $\beta$ using ``fit2'' and ``fit6'' of Table~\ref{tab:fit_many_w0}.}
  \label{fig:w0_versus_beta}
\end{figure}

\begin{figure}[H]
  \centering
  \hspace{1.5cm}
  \includegraphics[width=0.75\textwidth]{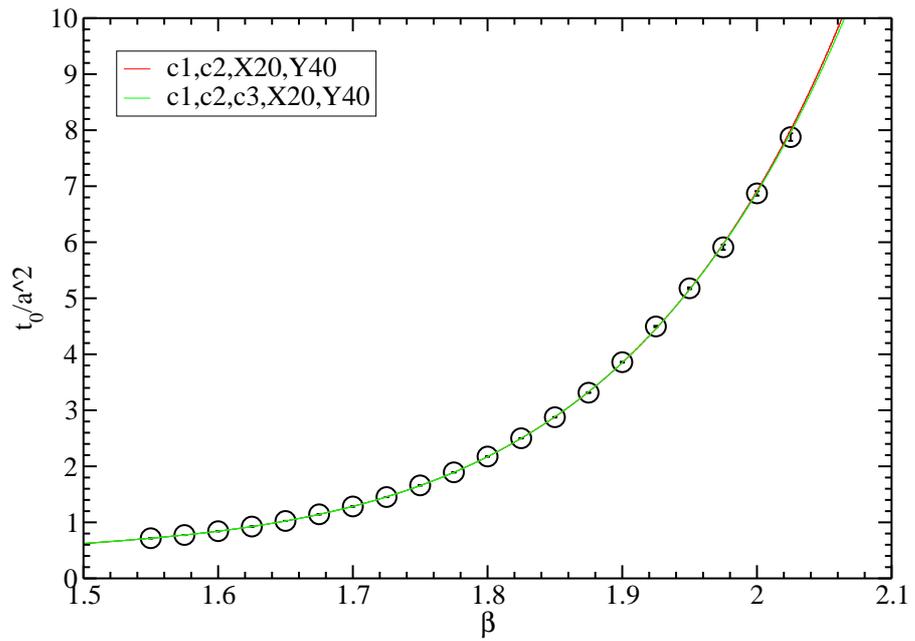}
  \caption{Interpolation of $t_0$ versus $\beta$ using ``fit9'' and ``fit10'' of Table~\ref{tab:fit_many_t0}.}
  \label{fig:t0_versus_beta}
\end{figure}

In Figure~\ref{fig:w0_versus_beta} we plot the scale $w_0$ as given in Table~\ref{tab:w0values} together
with the fits labelled as ``fit2'' and ``fit6'' in Table~\ref{tab:fit_many_w0}; overall the quality of the fit looks pretty decent. 
In Figure~\ref{fig:t0_versus_beta} we plot instead the scale $t_0$ of Table~\ref{tab:w0values} together 
with the best two fits labelled as ``fit9'' and ``fit10'' in Table~\ref{tab:fit_many_t0}. As final result we consider the average
of the two values, coming from the two fits, and as error the sum of the two statistical errors and
a systematic one which comes from the difference between the two values. 

\subsection{Final results of the lattice \texorpdfstring{$\beta$-function}{beta-function}}
\label{sec:finalbeta}

\begin{figure}[H]
  \centering
  \hspace{1.5cm}
  \includegraphics[width=0.75\textwidth]{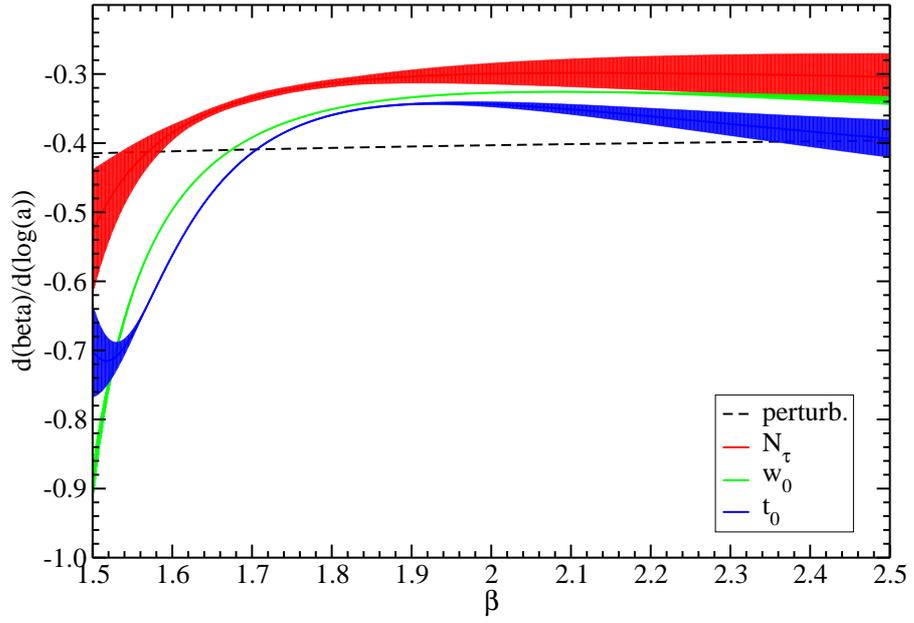}
  \caption{Plot $d\beta/d(\log(a))$.}
  \label{fig:dbetaoverlog}
\end{figure}

\begin{figure}[H]
  \centering
  \hspace{1.5cm}
  \includegraphics[width=0.75\textwidth]{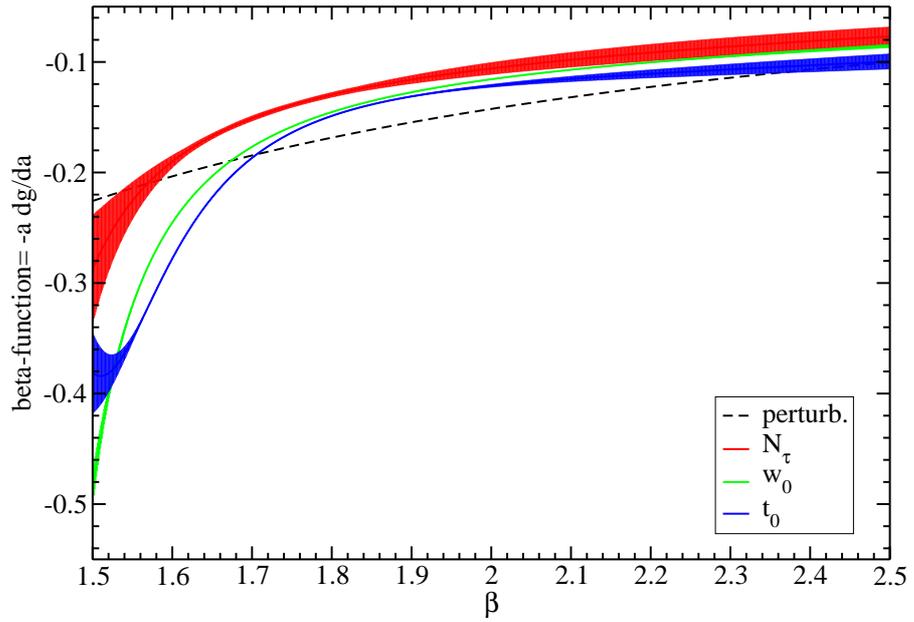}
  \caption{Plot $\beta$-function.}
  \label{fig:betafunction}
\end{figure}

In Figure~\ref{fig:dbetaoverlog} we plot the determination of ${\partial \beta}/{\partial \log(a)}$
and in Figure~\ref{fig:betafunction} that of the $\beta$-function for our three observables.
The perturbative dashed line present in the plots has been determined considering only
Eq.~(\ref{eq:fperturb}).
We have three different results which are not compatible with each other at low $\beta$, due to discretisation effects, which increase when $\beta$ decreases, and 
that are not universal, see Ref.~\cite{Allton:1996kr}.

\begin{figure}[H]
  \centering
  \hspace{1.5cm}
  \includegraphics[width=0.75\textwidth]{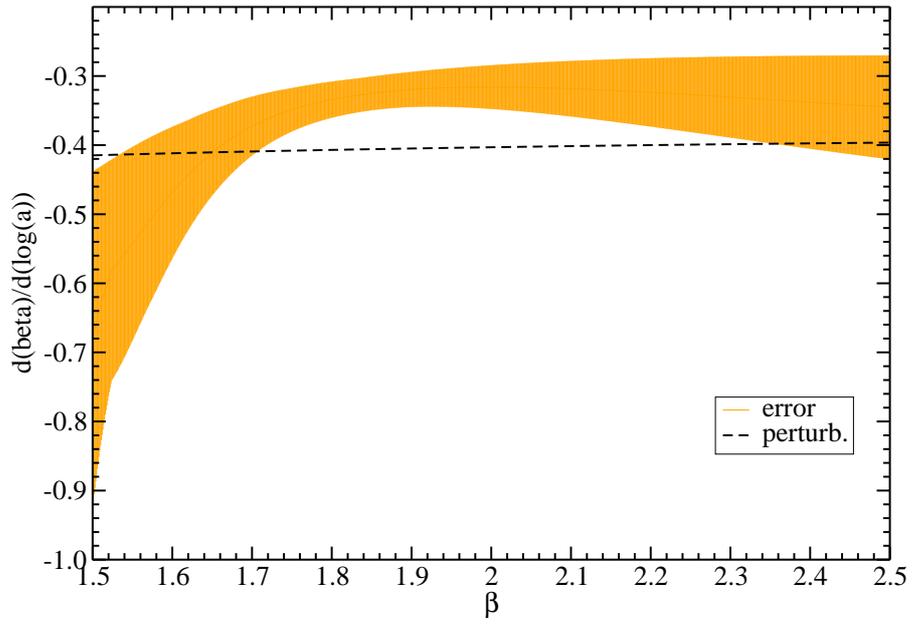}
  \caption{Plot $d\beta/d(\log(a))$: final uncertainty.}
  \label{fig:dbetaoverlog_err}
\end{figure}

\begin{figure}[H]
  \centering
  \hspace{1.5cm}
  \includegraphics[width=0.75\textwidth]{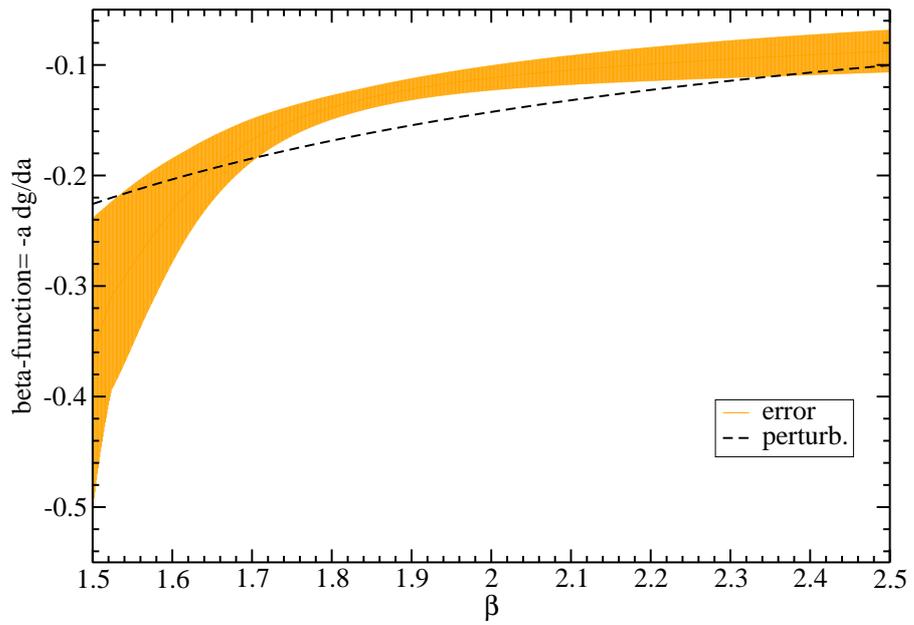}
  \caption{Plot $\beta$-function: final uncertainty.}
  \label{fig:betafunction_err}
\end{figure}

In this work the $\beta$-function and its error are safely defined by considering a combined final uncertainty that will arise mainly from the difference of the three possible observables used to set the scale. The final results is presented in Figure~\ref{fig:dbetaoverlog_err} and in Figure~\ref{fig:betafunction_err}. 
In any case, it is worth to consider what is the impact of such large discrepancy induced by violation of the scaling of the physical observables at low $\beta$. From Figure~\ref{fig:betafunction}, the stronger difference between the various $\beta$-functions appears for $\beta \lesssim 1.8$, that corresponds to $T/T_c \lesssim 1$ for $N_\tau=5$ (see Figure~\ref{fig:TvsTc_versus_beta}), {\itshape i.e.} the largest uncertainties of the $\beta$-function are in the confined phase, where thermodynamical quantities, such as pressure and energy density, are usually very small.

A method to avoid at least part of the previous uncertainties in the calculation of thermodynamical quantities must provide a direct definition of the energy-momentum tensor and of its renormalisation on the lattice, some work in this direction has been presented for instance in Ref.~\cite{Asakawa:2013laa}, based on the gradient flow, 
or in Ref.~\cite{Giusti:2016iqr}, based on a formulation of the thermal theory in a moving reference frame. In any case, discretisation errors are unavoidable in any lattice numerical simulations and will appear in the determination of the equation of state both in the ordinate for renormalised quantities, and in the abscissa as uncertainties in the definition of the physical temperature. As we will show in the following sections, the use of the Symanzik improved action is crucial to suppress lattice discretisation errors without requiring at the same time a demanding computational cost.


\section{Thermodynamics}
\label{sec:themodynamics}

The thermodynamic quantities we are interested in are the pressure $p$,
the energy density $\epsilon$, the trace anomaly $\Delta$ and 
the entropy density $s$. They are defined from the partition function
\beq
Z(T,V)=\int DU e^{-\beta S} \ ,
\eeq
according to the relations
\bea
f & = & -\frac{T}{V} \log Z \ ,  \\
p & = & T \left. \frac{\partial \log Z}{\partial V} \right|_T  \ ,  \\
\epsilon & = & \frac{T^2}{V} \left. \frac{\partial \log Z}{\partial T} 
\right|_V  \ ,  \\
\Delta &=& T^5 \left. \frac{\partial}{\partial T} \left( \frac{p}{T^4} \right)
\right|_V   =  \epsilon - 3p \ ,  \\
s &=&  \frac{\epsilon-f}{T} \,,
\eea
being $f$ the Helmholtz free energy. The pressure is determined using the integral method, see Ref.~\cite{Engels:1990vr}:
\beq
\frac{p}{T^4}= \frac{1}{T^4} \int_{\beta_0}^{\beta} d\beta \left[ \langle
\mathcal{S} \rangle_0 -\langle \mathcal{S} \rangle_T \right] \ , 
\label{eq:pressure}
\eeq
where we have introduced the action density $\langle \mathcal{S} \rangle = (T/V) \langle S \rangle$. Note the existence of a reference $\beta_0$ which should correspond 
to a sufficiently small temperature where the pressure can be safely assumed to be zero (for a different approach see Ref.~\cite{Endrodi:2007tq}).
The trace anomaly is given by
\beq
\frac{\Delta}{T^4}=\frac{1}{T^4}  \left[ \langle \mathcal{S} \rangle_0 
-\langle \mathcal{S} \rangle_T \right] 
\frac{\partial \beta}{\partial \log{a}} \ .
\eeq
The other two quantities are determined as a function of the pressure and of the trace anomaly:
\bea
\frac{\epsilon}{T^4}&=&\frac{\Delta+3 p}{T^4} \ , \\
\frac{s}{T^3}       &=&\frac{\Delta+4 p}{T^4} \ .
\eea
The Stephan Boltzmann (SB) limit for these quantities is known to be equal to
\bea
\Delta&=&0 \ , \label{eq:delta} \\
\frac{p}{T^4}&=&\frac{\pi^2}{45}(N_c^2-1) \ , \label{eq:press}  \\
\frac{\epsilon}{T^4}&=&\frac{\pi^2}{15}(N_c^2-1) \ , \label{eq:epsilon}   \\
\frac{s}{T^3}&=&\frac{4 \pi^2}{45}(N_c^2-1) \ . \label{eq:entropy}  
\eea
We will use these relations to normalise our final results.

Note that Eqs.~(\ref{eq:delta})--(\ref{eq:entropy})  are not taking into account the correction to the canonical partition function 
which is proportional to $\ln{(N_s/N_\tau)}$, see Refs.~\cite{Gliozzi:2007jh, Panero:2008mg}.

\subsection{Simulations}
\label{sub:simulations}

Two different values of the temporal extension of the lattice, $N_\tau=4$ and  $N_\tau=5$, have been used for the simulations at non-zero temperature, while we have fixed an aspect ratio of $N_s/N_\tau=6$, see Table~\ref{tab:summary_simulations_finite_T}. The remarkable result is that the second lattice, with temporal extension $N_\tau=5$, turned out to be already close to the continuum limit, with a small correction with respect to the results coming from the lattice with $N_\tau=4$.

For $N_\tau=4$, we have measured the action density and the Polyakov loop in the interval $1.550 < \beta < 2.165$ and for $N_\tau=5$ in the interval $1.655 < \beta < 2.330$; in both cases the measurements were done every $\Delta\beta=0.005$.

Note that finite volume effects depend on the ratio $\hat\xi/N_s$, where $\hat\xi$ is the correlation length: far from the critical $\beta$ this ratio goes to zero and there are small finite volume effects. On the contrary, close to the deconfinement phase transition, the correlation length will diverge for a second order phase transition, as in the case of the SU(2) Yang-Mills theory \cite{Pelissetto:2000ek}. Therefore one can set a decreasing aspect ratio increasing the distance from the critical $\beta$ and our value of $N_s/N_\tau=6$ is pretty arbitrary, tuned to control finite volume effects near the critical temperature.

\begin{table}[htbp]
  \begin{center}
    \begin{tabular}{|l|l|l|l|l|l|}
      \hline
      $N_\tau$ & $N_s$ & $N_s/N_\tau$    & $\beta_{min} - \beta_{max}$ & Confs   \\
      \hline
      4       & 16                  & 4              & 1.625 - 1.715 & 208000 \\  
              & 20                  & 5              & 1.625 - 1.715 & 208000 \\
              & 24                  & 6              & 1.550 - 2.165 & 100000 \\
              & 28                  & 7              & 1.625 - 1.715 & 76000  \\
              & 32                  & 8              & 1.625 - 1.715 & 51000  \\
      \hline
      5       & 20                  & 4              & 1.690 - 1.790 & 1530000 \\  
              & 25                  & 5              & 1.690 - 1.790 &  875000 \\  
              & 30                  & 6              & 1.655 - 2.330 &  400000 \\  
              & 35                  & 7              & 1.690 - 1.790 &  256000 \\  
              & 40                  & 8              & 1.690 - 1.790 &  240000 \\  
      \hline
    \end{tabular}
  \end{center}
  \caption{Summary of the simulations employed at finite temperature. The interval used to span the $\beta$ interval is always $\Delta\beta=0.005$.}
  \label{tab:summary_simulations_finite_T}
\end{table}

\begin{figure}[H]
  \centering
  \hspace{1.5cm}
  \includegraphics[width=0.65\textwidth]{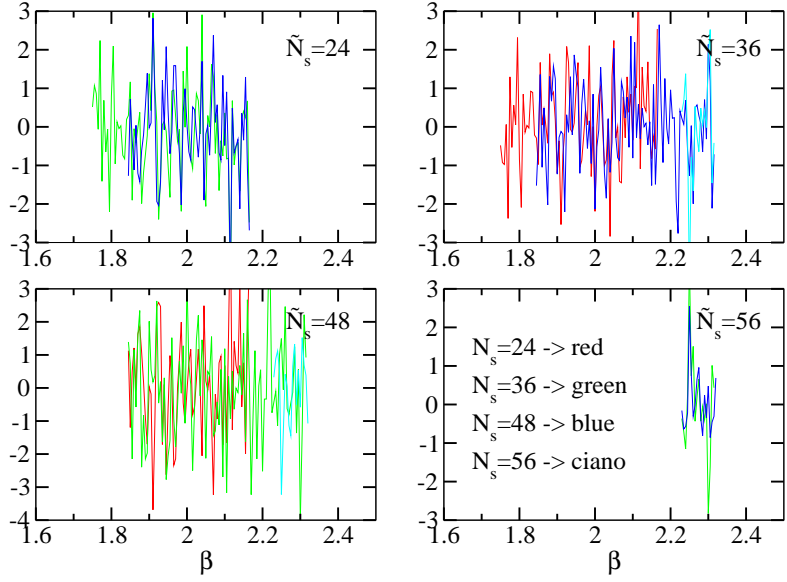}
  \caption{In this plot we show the finite volume effects of our simulations at zero temperature. We plot
  the quantity $[P(N_s)-P(\tilde{N}_s)]/ \Delta{ P(\tilde{N}_s) }$, where $P(N_s)$ is the value of the spatial plaquette
  measured at the spatial volume $N_s^4$ and $\Delta{P}$ is its error. Comparing the results from different volumes, 
  in the same range of $\beta$, we can see that the difference is always smaller than three standard deviations.
  }
  \label{fig:finitevolumeeffects}
\end{figure}

In Table~\ref{tab:summary_simulations_zero_T} we show the details of our simulations at zero temperature, generated in order to perform the subtraction of the zero temperature expectation value of the action density. We show also
the ``residual'' non-zero temperature in each case. Its relevance and that of the finite volume effects can be seen in 
Figure~\ref{fig:finitevolumeeffects} where we have compared the value of the spatial plaquette for different volumes
along the entire range of $\beta$ where we have used our data. We have plotted the ratio:
\beq
[P(N_s)-P(\tilde{N}_s)]/ \Delta{ P(\tilde{N}_s) } \ , 
\eeq
where $P(N_s)$ is the value of the spatial plaquette measured at the spatial volume 
$N_s^4$ and $\Delta{P}$ is its error. $\tilde{N}_s$ labels the value with respect to
we are comparing the data and it is fixed in each single plot. 
From this figure it is clear that the ``residual'' temperature and the finite volume effects are always smaller than
the statistical fluctuation of our measurements.
\begin{table}[htbp]
  \begin{center}
    \begin{tabular}{|l|l|l|l|l|l|}
      \hline
      $N_\tau=N_s$ & $\beta$-range             & $T/T_c$-range    & Confs   \\
      \hline
      24                        & 1.550 - 1.745             & 0.11 - 0.19        & 20000  \\
      36                        & 1.750 - 1.840             & 0.13 - 0.17        & 4000  \\
      48                        & 1.845 - 2.225             & 0.13 - 0.46        & 2700  \\
      56                        & 2.230 - 2.260             & 0.40 - 0.45        & 1210  \\
      \hline
    \end{tabular}
  \end{center}
  \caption{Summary of the simulations used at zero temperature. The interval used to span the $\beta$ interval is always $\Delta\beta=0.005$.
    The range in $T/T_c$ is the ``residual'' non-zero temperature of the system determined from Figure~\ref{fig:Nt_versus_beta}. }
  \label{tab:summary_simulations_zero_T}
\end{table}

The error in the determination of our observables depends on the error on the
action density $N_\tau^4 \left[ \langle \mathcal{S} \rangle_0 -\langle \mathcal{S} \rangle_T \right]$. 
It is possible to show that, for a noninteracting theory, this error is proportional to
$N_\tau^{3.5}/N_s^{1.5}$.

This relation explains why the number of configurations that must be used in order to get a reasonably small statistical error increases hugely moving toward the continuum limit $N_\tau \rightarrow \infty$. In our case going from $N_\tau=4$ to $N_\tau=5$ required already an increase by roughly a factor five of the computational cost. 

\subsection{Action density and finite size scaling}

Since the SU(2) gauge theory is characterised by a 
second order phase transition, see Ref.~\cite{Pelissetto:2000ek},
strong finite volume effects are present around the critical temperature. 
It is therefore necessary 
to simulate different volumes to extrapolate
to the infinite volume limit. We have several ensembles with five different
volumes both at $N_\tau=4$, with volumes $(16,20,24,28,32)$, and 
at $N_\tau=5$ with volumes $(20,25,30,35,40)$, see Table~\ref{tab:summary_simulations_finite_T}.

Close to the critical temperature, the infinite volume limit has been extrapolated using the finite-scaling approach.
The action density $\langle \mathcal{S} \rangle$ is a 
lattice  operator which, under  Svetitsky-Yaffe conjecture,
is mapped into the energy operator of a statistical model, as 
shown in Ref.~\cite{Gliozzi:1997yc}.
The scaling behaviour for $\langle \mathcal{S} \rangle$ is given by:
\beq
\langle \mathcal{S} \rangle_L(t) = \langle \mathcal{S} \rangle_\infty(t) +
L^{1/\nu-d} Q_\mathcal{S}(t L^{1/\nu}) \ ,
\label{FSscalinggeenral}
\eeq
where $L$ is the spatial extension and $Q_\mathcal{S}$ is the scaling function for  
$\langle \mathcal{S} \rangle$.
At $t=0$ we can therefore extrapolate the action density to the infinite volume 
limit following the ansatz (see also Refs.~\cite{Papa:2002gt,Engels:1989fz}):
\beq
\langle \mathcal{S} \rangle_L = \langle \mathcal{S} \rangle_\infty + A
L^{1/\nu-d}  \ .
\label{FSscaling}
\eeq
The critical indexes for SU(2) in $4d$ are those of the Ising model 
in $3d$ (see section 3.2.1 of Ref.~\cite{Pelissetto:2000ek}):
\bea
\nu&=&0.6301(4) \ , \\
\gamma&=&1.2372(5)  \ ,    \\
\beta&=&0.3265(3) \ .
\eea
We have verified that the action density $\mathcal{S}$ 
is affected by finite size effects, non compatible with the statistical errors, 
in the interval $T/T_c\in [0.985,1.005]$. 

Since Eq.~(\ref{FSscaling}) is valid only at the critical point,
we have tried to expand perturbatively Eq.~(\ref{FSscalinggeenral})
for $t \ne 0$, see 
Refs.~\cite{Fisher:1972zza, Privman:1984zz, Blote:1995zik, Barber:1983dl},
to extrapolate the results to infinite volume. Unfortunately, 
we have only results for five volumes that are not enough to allow 
a stable and reliable numerical extrapolation. 
Therefore, we follow the ansatz of Eq.~(\ref{FSscaling}) in the entire critical
region $T/T_c\in [0.985,1.005]$. Anyway, to take into account  
the systematic error due to the sloppy infinite 
volume extrapolation, we verified the scaling of
this quantity, plotting $(\langle \mathcal{S} \rangle_L-
\langle \mathcal{S}\rangle_\infty) L^{1/\nu-d}$ vs 
$t L^{1/\nu}$. Because the five curves were not compatible with each other, 
we increased arbitrary, in the critical region, the statistical error of 
$\langle \mathcal{S}\rangle_\infty$ until the five curves were made compatible.
At the end we tripled the statistical error at $N_\tau=4$ and we doubled at $N_\tau=5$.
Thanks to this procedure, the value of $\langle \mathcal{S}\rangle_\infty$ and its error
are determined in a way which should be able to correctly estimate the presence
of the systematic error.

\begin{figure}[H]
  \centering
  \hspace{1.5cm}
  \includegraphics[width=0.65\textwidth]{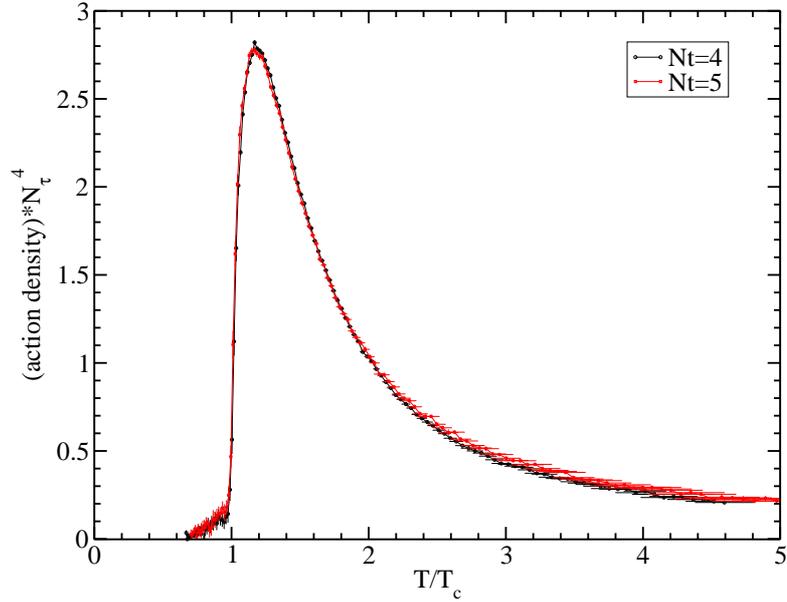}
  \caption{Action density $\left[ \langle \mathcal{S} 
\rangle_0 -\langle \mathcal{S} \rangle_T \right] N_\tau^4 $. Note that results 
obtained at two different lattice spacings are compatible.}
  \label{fig:diffactiondensity}
\end{figure}

The final value of the action density, normalised to the $T=0$ value,
is plotted in Figure~\ref{fig:diffactiondensity}. It is interesting
to note that the results obtained at $N_\tau=4$ and at $N_\tau=5$,
which correspond to a smaller lattice spacing, are compatible.
Discretisation errors smaller than all the other errors are possible because we are using an improved action which
brings our results already close to the continuum limit. 
We do not need therefore in our analysis to introduce any correction 
term $R_I(N_\tau)$, as done for example in Ref.~\cite{Panero:2009tv}.
As a matter of fact, as discussed in Ref.~\cite{Beinlich:1995ik},
the expected correction for our action is, at $N_\tau=4$, of the order of 
$1.35\%$ which is smaller than our statistical error.

\begin{figure}[H]
  \centering
  \hspace{1.5cm}
  \includegraphics[width=0.75\textwidth]{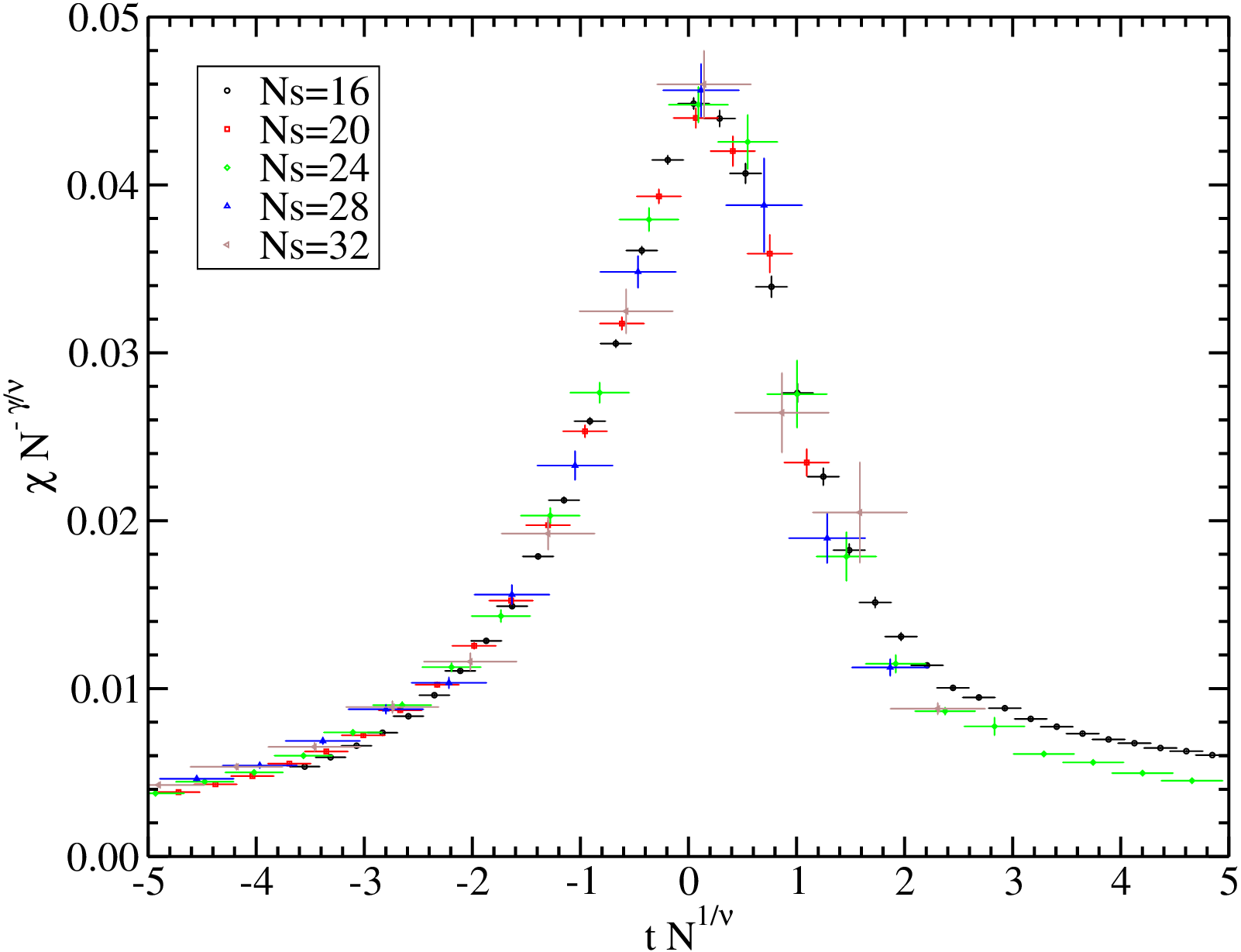}
  \caption{Scaling of the susceptibility of the Polyakov loop for $N_\tau=4$.}
  \label{fig:scalingsusceptpolykovNt4}
\end{figure}

\begin{figure}[H]
  \centering
  \hspace{1.5cm}
  \includegraphics[width=0.75\textwidth]{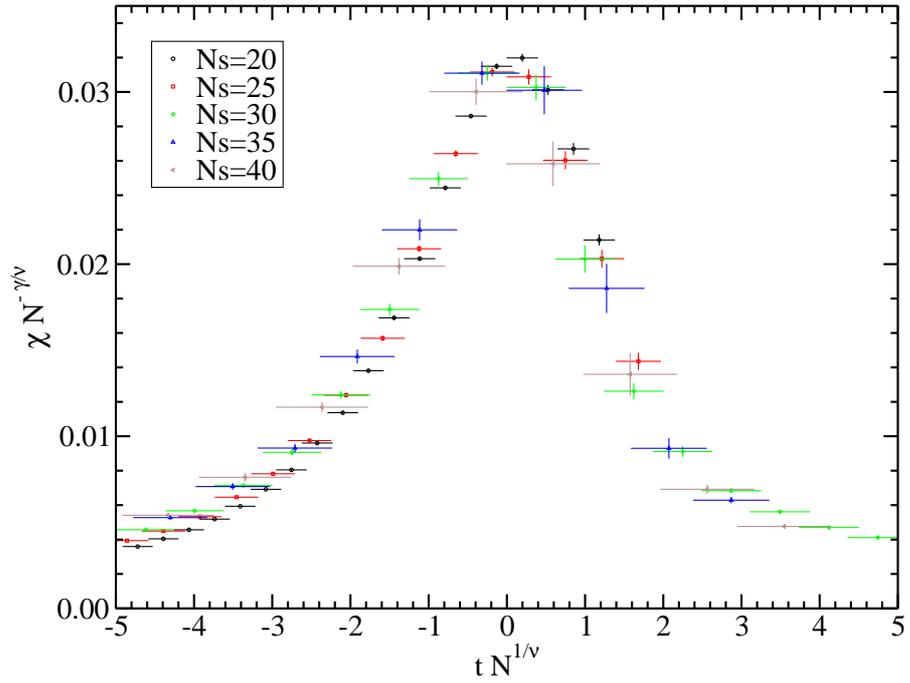}
  \caption{Scaling of the susceptibility of the Polyakov loop for $N_\tau=5$.}
  \label{fig:scalingsusceptpolykovNt5}
\end{figure}

It is interesting to look at the scaling of the susceptibility
of the Polyakov loop, which is given by a scaling function $Q_\chi$ without a constant term: 
\beq
\chi(t)=L^{\gamma/\nu} Q_\chi(t L^{1/\nu}) \ .
\eeq
We show the results in Figure~\ref{fig:scalingsusceptpolykovNt4} and
in Figure~\ref{fig:scalingsusceptpolykovNt5}  (see for comparison 
Ref.~\cite{Engels:1989fz}). Clearly the susceptibility follows the scaling relation in a very wide range of $t$.

\subsection{Thermodynamic results}

Using the relations introduced at the beginning of Sec.~\ref{sec:themodynamics}, the
action density plotted in Figure~\ref{fig:diffactiondensity} and 
the derivative of the $\beta$-function of Figure~\ref{fig:dbetaoverlog_err}, 
we can now determine all the other thermodynamic observables.

\begin{figure}[H]
  \centering
  \hspace{1.5cm}
  \includegraphics[width=0.75\textwidth]{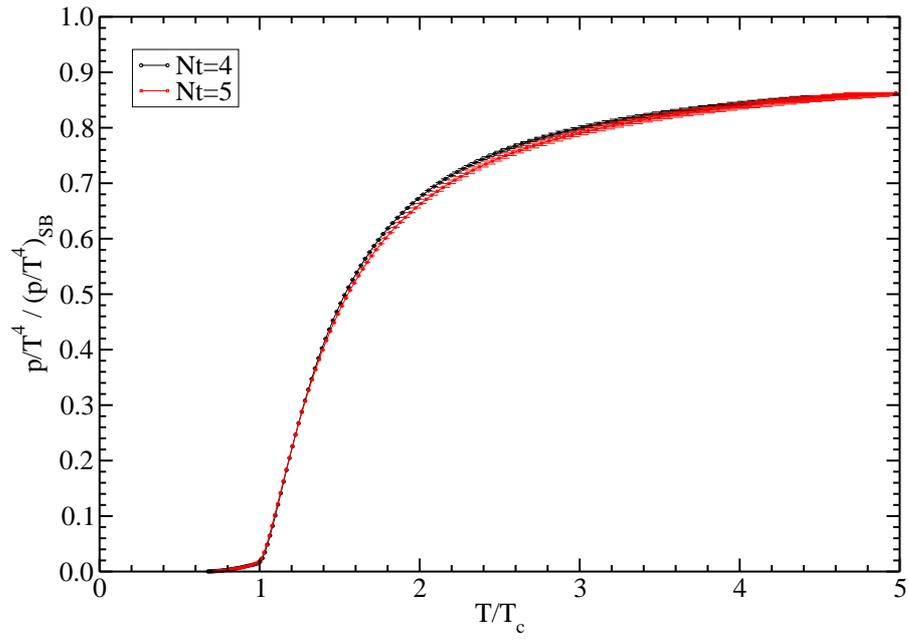}
  \caption{Pressure normalised to the SB limit.}
  \label{fig:pressurenorm}
\end{figure}

\begin{figure}[H]
  \centering
  \hspace{1.5cm}
  \includegraphics[width=0.75\textwidth]{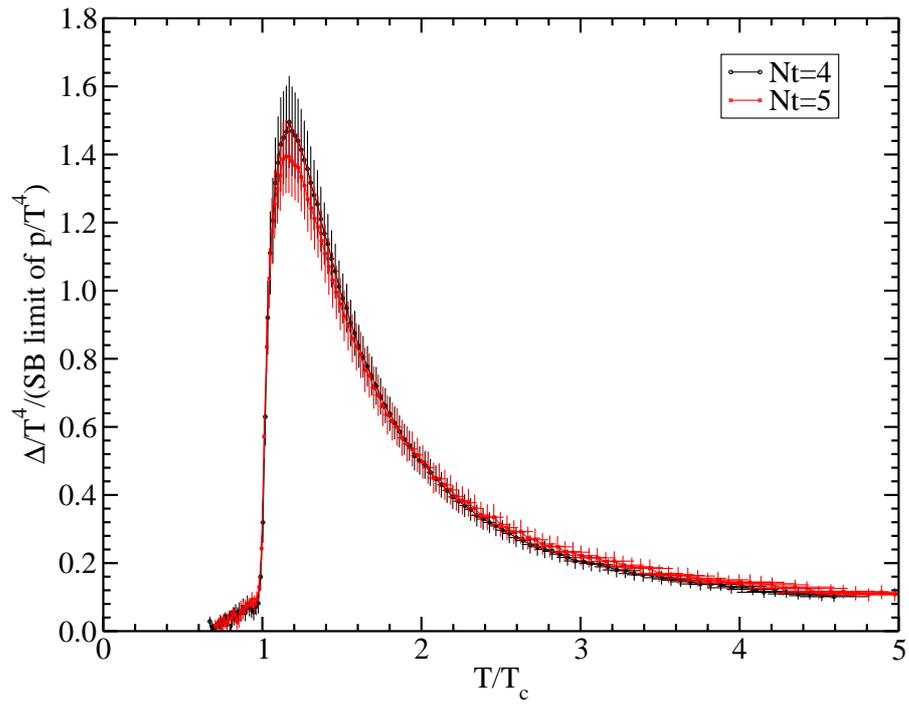}
  \caption{Trace anomaly normalised to the SB limit of the pressure.}
  \label{fig:traceanomalynorm}
\end{figure}

\begin{figure}[H]
  \centering
  \hspace{1.5cm}
  \includegraphics[width=0.75\textwidth]{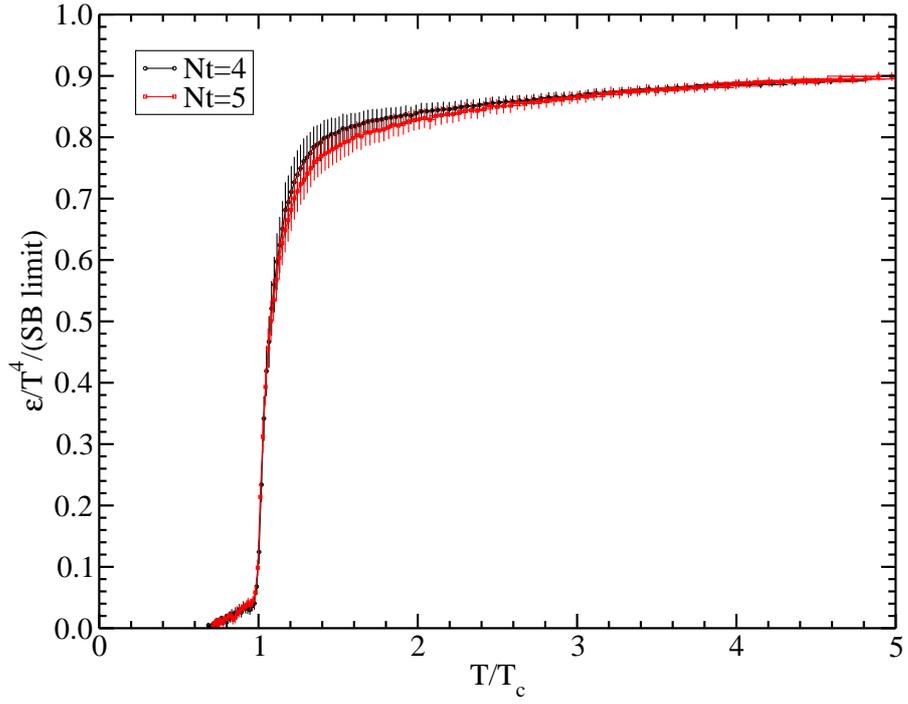}
  \caption{Energy density $\epsilon$ normalised to the SB limit.}
  \label{fig:energydensitynorm}
\end{figure}

\begin{figure}[H]
  \centering
  \hspace{1.5cm}
  \includegraphics[width=0.75\textwidth]{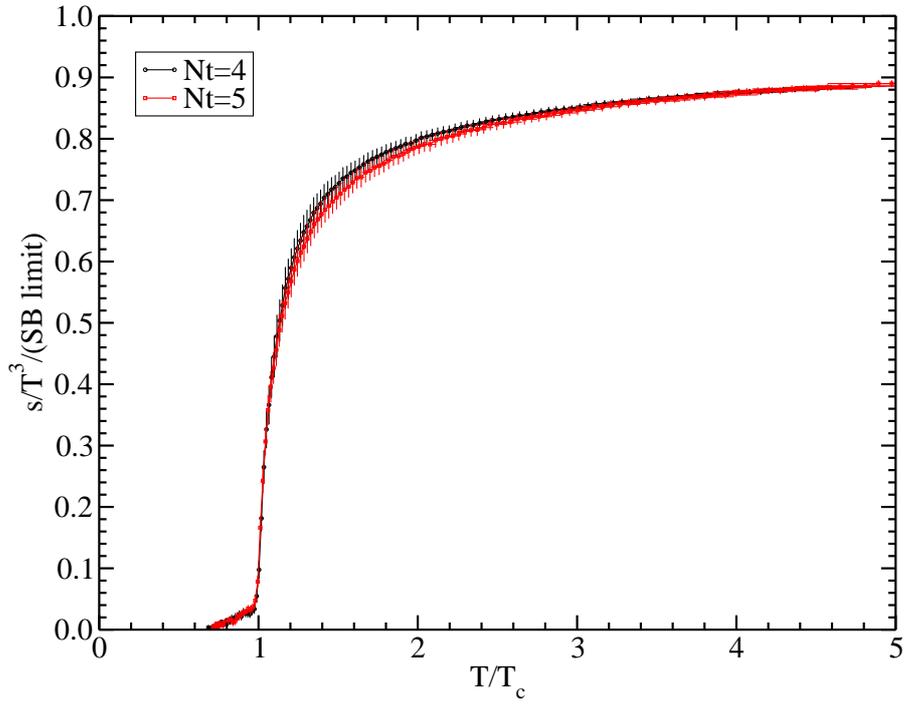}
  \caption{Entropy density $s$ normalised to the SB limit.}
  \label{fig:entropynorm}
\end{figure}

The pressure, normalised to its SB value, is plotted in Figure~\ref{fig:pressurenorm}. 
In Figure~\ref{fig:traceanomalynorm}, we plot the trace anomaly normalised to the SB 
value of the pressure (as has been done in Ref.~\cite{Panero:2009tv}).
The SB normalised energy and entropy densities can be found respectively 
in Figure~\ref{fig:energydensitynorm} and in Figure~\ref{fig:entropynorm}.

As can be seen all our observables reach a value around $90\%$ of the SB limit 
at $T/T_c=5$ and the results obtained at $N_\tau=4$ and $N_\tau=5$ are 
compatible, confirming that the discretisation effects are under control.


\section{Comparison with other works}
\label{sec:comparison}

\begin{figure}[H]
  \centering
  \hspace{1.5cm}
  \includegraphics[width=0.75\textwidth]{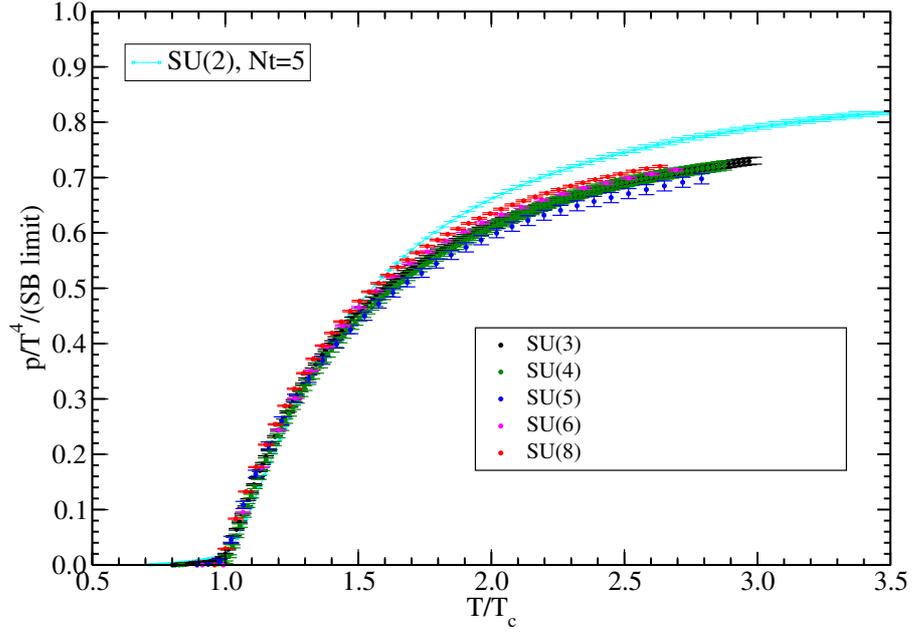}
  \caption{Comparison of the pressure in the deconfined phase, measured in this work, {\itshape i.e.} SU(2),
    with the determinations in SU(N$_c$) of Ref.~\cite{Panero:2009tv}.}
  \label{fig:comparison_deconf_press}
\end{figure}

\begin{figure}[H]
  \centering
  \hspace{1.5cm}
  \includegraphics[width=0.75\textwidth]{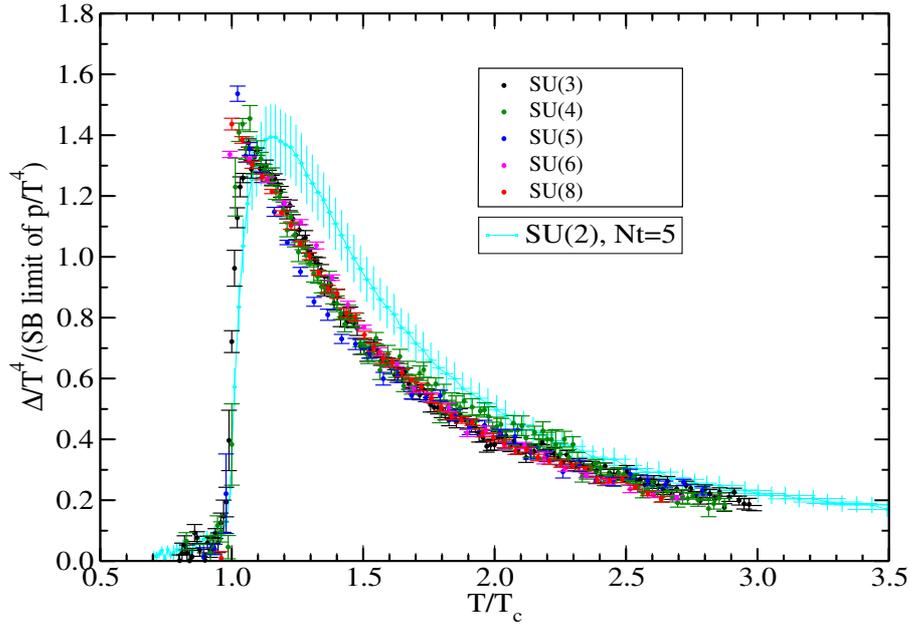}
  \caption{Comparison of the trace anomaly in the deconfined phase, measured in this work, {\itshape i.e.} SU(2),
    with the determinations in SU(N$_c$) of Ref.~\cite{Panero:2009tv}.}
  \label{fig:comparison_deconf_anomaly}
\end{figure}

It is interesting to compare our results, in the deconfined phase, with 
those of  Ref.~\cite{Panero:2009tv} where results for SU(N$_c$), and 
$N_c \geq 3$ have been considered. 
Note that in that work was used only the standard Wilson action and only 
one volume, therefore we expect that both discretisation and finite volume 
effects are present.
All thermodynamical observables we have considered reach the SB limit 
quicker than in Yang-Mills theories with $N_c \geq 3$. For
example, in Figure~\ref{fig:comparison_deconf_press}, 
the value of the pressure at $T/T_c=3.0$ is $\sim 10\%$ higher. 
The difference can be better appreciated comparing directly the
trace anomaly, see Figure~\ref{fig:comparison_deconf_anomaly}. In this
case a huge difference appears around 1.5$T_c$ and the value is 
always above the $N_c \geq 3$ case.

\begin{figure}[H]
  \centering
  \hspace{1.5cm}
  \includegraphics[width=0.75\textwidth]{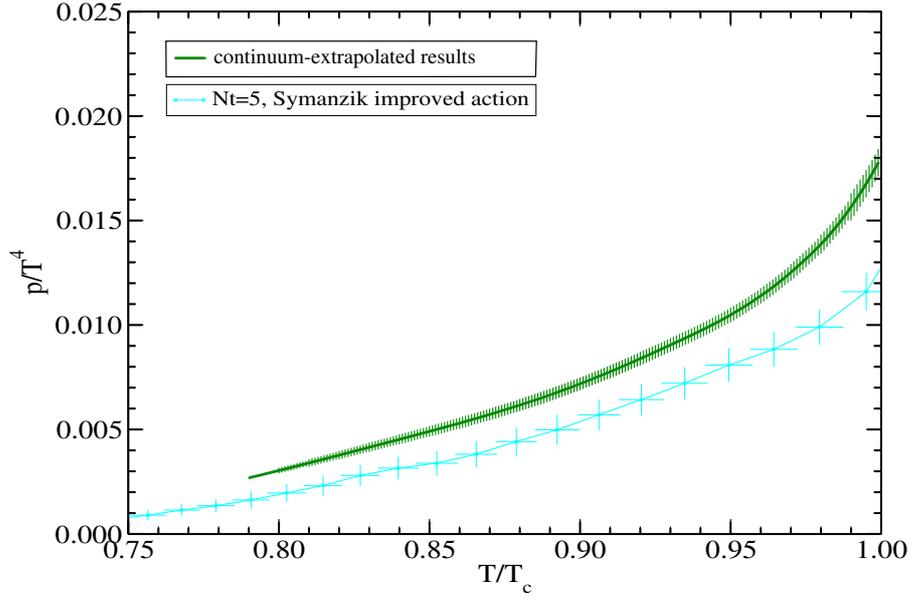}
  \caption{Comparison of the pressure in the confined phase, measured in this work (cyan line),
    with data taken from Ref.~\cite{Alba:2016fku}(green line).}
  \label{fig:comparison_conf_press}
\end{figure}

Moreover, we can compare our results, in the confined phase, with those
published in Refs.~\cite{Caselle:2015tza, Alba:2016fku} where they simulate
SU(2) pure gauge theory but, also in this case, using  
the standard Wilson action at fixed volume. However, they simulate
different values of $N_\tau$, ranging from 5 to 10.
In Figure~\ref{fig:comparison_conf_press} we compare the pressure
with their continuum extrapolated results. Our values are always smaller 
and, close to the critical point, the value is about $1.4$ times smaller.
This difference can be explained by finite volume effects,
that, as we have already seen, have a strong effect in this theory.

\begin{figure}[H]
  \centering
  \hspace{0.0cm}
  \includegraphics[width=0.85\textwidth]{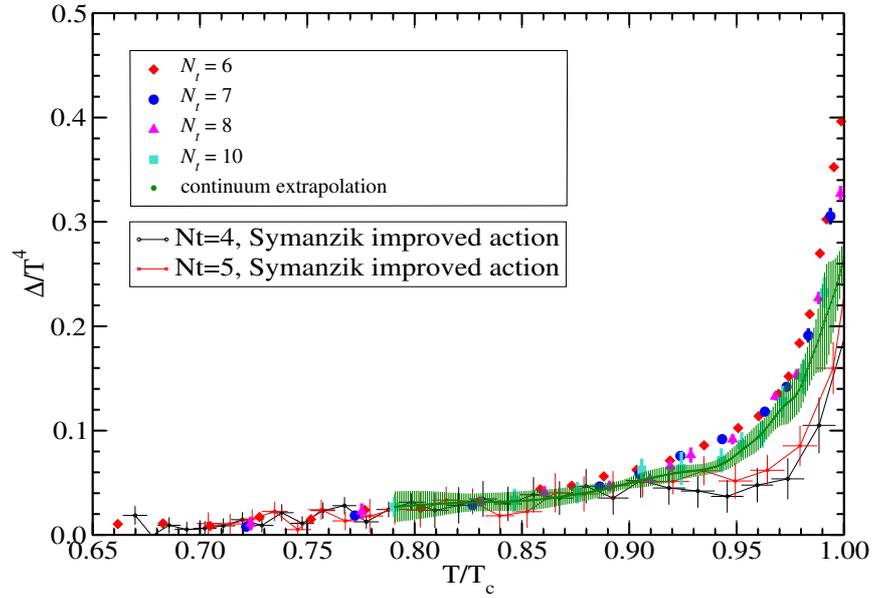}
  \caption{Comparison of the trace anomaly in the confined phase, measured in this work,
    with data taken from Ref.~\cite{Alba:2016fku}.}
  \label{fig:comparison_conf_anomaly}
\end{figure}

The results of the pressure can be affected by the choice of the 
reference point $\beta_0$, see Eq.~(\ref{eq:pressure}), therefore we compare also the trace anomaly directly with 
our results in Figure~\ref{fig:comparison_conf_anomaly}.
Also in this case we can see a clear difference in the range 
$0.9 \lesssim  T/T_c \lesssim 1.0$. In this case, the discrepancy could be given by 
the different choice of the $\beta$-function.

The observation that the trace anomaly falls off as $1/T^2$ above $T_c$
lead to the development of many phenomenological models, 
see Refs. \cite{Meisinger:2001cq, Megias:2005ve, Pisarski:2006hz}.
It is therefore interesting to compare SU(2) with 
previous studies where $N_c \geq 3$ was considered.  

\begin{figure}[H]
  \centering
  \hspace{0.0cm}
  \includegraphics[width=0.75\textwidth]{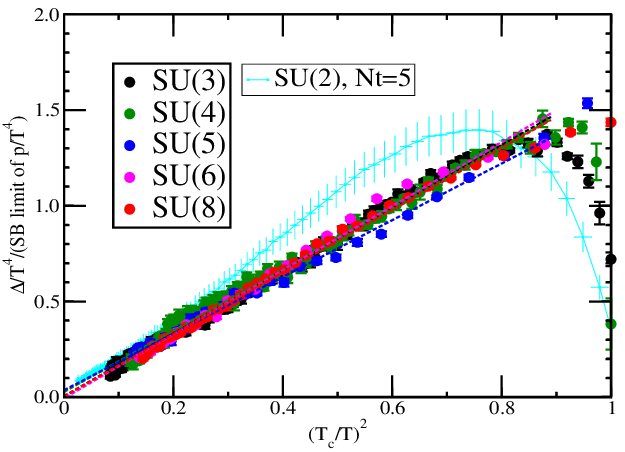}
  \caption{Comparison of the trace anomaly in the deconfined phase with Ref.~\cite{Panero:2009tv} .}
  \label{fig:comparison_traceanomaly1}
\end{figure}

\begin{figure}[H]
  \centering
  \hspace{0.0cm}
  \includegraphics[width=0.75\textwidth]{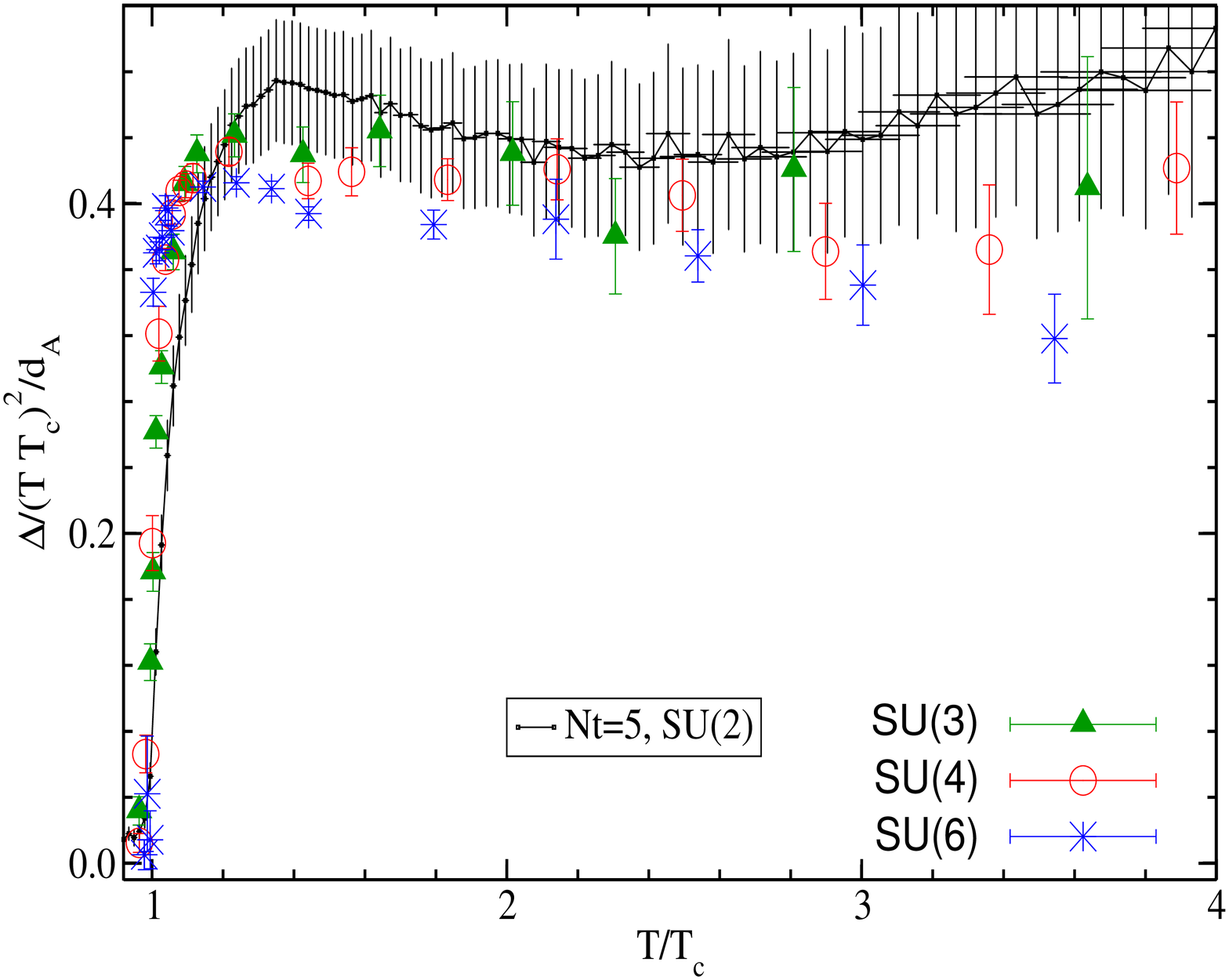}
  \caption{Comparison of the trace anomaly in the deconfined phase with Ref.~\cite{Datta:2010sq}.}
  \label{fig:comparison_traceanomaly2}
\end{figure}

In Figure~\ref{fig:comparison_traceanomaly1} we compare our results with Ref.~\cite{Panero:2009tv}.
We plot the quantity $\Delta/T^2$ versus $(T_c/T)^2$ to see whether there exists a region with a linear behaviour. 
The figure suggests that SU(2) is compatible with the other theories only for temperature 
above $\approx 2$ $T/T_c$. Otherwise, for temperature from $T/T_c$ until $2 T/T_c$ the values for SU(2) are larger
and not compatible with the others  within the errors.
The difference could be guessed already clearly from Figure~\ref{fig:comparison_deconf_anomaly}.
A real difference between SU(2) and SU($N_c$) Yang-Mills theory could be claimed only after all systematic errors are carefully taken into account. 
Residual finite volume effects in our results, 
different $\beta$-functions, missing of continuum limit in the results 
of Ref.~\cite{Panero:2009tv} could be at the origin of the discrepancy that we observe.  
However, in Figure~\ref{fig:comparison_traceanomaly2}
we plot the quantity $\Delta/(T^2 T_c^2 d_A)$ ($d_A=N_c^2-1$) and our results 
show a better compatibility with those of Ref.~\cite{Datta:2010sq}, 
where both the thermodynamic and to the continuum limit have been extrapolated. 
We can therefore state that our results do not exclude the possibility that 
also for SU(2) the trace anomaly has a $1/T^2$ behaviour, even if further 
simulations are necessary.

It is clear, by the examples considered, how much the use of an 
improved action is important to study the thermodynamic properties of a QCD-like theory, in particular
when analytical determinations are compared to lattice results, as for the
holographic model in Ref.~\cite{Panero:2009tv}, for the effective bosonic
string model in Ref.~\cite{Caselle:2015tza}, or for the
hadron-resonance-gas model in Ref.~\cite{Alba:2016fku}.


\section{Conclusions}
\label{sec:conclusion}

In this paper we have presented our results concerning the thermodynamics
of SU(2) pure gauge theory. This is the first work, after almost twenty years,
where a systematic study of the equation of state of this theory has been performed.
The SU(2) Yang-Mills theory can still be useful to compare and test some interesting models, which 
go from effective string descriptions to large $N_c$-limit results and from 
holografic models to quasi-particles descriptions. 
For our simulations we have used a Symanzik improved action so that our results,
already at $N_\tau=5$, are
compatible with the continuum limit within the statistical errors.

We have performed many simulations on different volumes near the deconfinement transition, to control finite volume effects that are significant for a theory with a second order phase transition. We extrapolated our results to the thermodynamic limit following the finite size-scaling relations.
We have determined non-perturbatively, employing three different methods,
the $\beta$-function and later we have determined the main thermodynamic observables for the equation of state.

Finally we have compared our results, both in the confined and deconfined phase, with
previous works, where clearly emerge the importance of using an improved action
in this kind of measurements, in particular when we want to compare lattice results 
with analytic models.


\section*{Acknowledgments}

We thank Prof. Paolo Castorina for having risen the problem of the lack
of updated data concerning the thermodynamics of SU(2) and for the
many discussions at the beginning of the work. We thank moreover
Dr Paolo Alba and Prof.~Robert~D.~Pisarski for their comments on the first
version of this paper.
SP acknowledges support from the Deutsche Forschungsgemeinschaft 
Grant No. SFB/TRR 55.
The calculations were carried out on the computer cluster PALMA of 
the University of Münster.


\newpage
\clearpage

\newpage

\newpage

\newpage

\newpage

\newpage

\newpage


\newpage


\newpage

\newpage

\newpage

\end{document}